\documentclass[a4paper,conference]{IEEEtran}
\usepackage{cite}      % Written by Donald Arseneau
\usepackage{graphicx}  % Written by David Carlisle and Sebastian Rahtz

\usepackage{subfigure} % Written by Steven Douglas Cochran
\usepackage{fancyheadings,aas_macros,bm,amsmath}
\begin{document}

\title{Paper Title }

\title{On shock capturing in SPH}

\author{\IEEEauthorblockN{Daniel J. Price}
\IEEEauthorblockA{School of Physics \& Astronomy \\
Monash University\\
Vic. 3800, Australia\\
daniel.price@monash.edu}
}

%\specialpapernotice{(Invited Paper)}

\maketitle

\begin{abstract}
For the past 20 years, our approach to shock capturing in smoothed particle hydrodynamics (SPH) has been to use artificial viscosity and conductivity terms supplemented by switches to control excess dissipation away from shocks \cite{monaghan97,morrismonaghan97}. This approach has been demonstrated to be superior to approximate Riemann solvers in a recent comparison \cite{puriramachandran14}. The Cullen \& Dehnen switch \cite{cullendehnen10} is regarded as the state of the art. But are we missing something? I will present a novel approach to shock capturing in SPH that utilises the philosophy of approximate Riemann solvers but provides a direct improvement on the ability to reduce excess dissipation away from shocks while preserving the fidelity of the shock itself.
\end{abstract}

\section{Introduction}
 What starts in fun, ends in tears. So hyperbolic equations always lead to trouble. Even the most innocent, small amplitude wave, if left to run to $t\to \infty$ according to the compressible Euler equations, will end up producing an infinite velocity gradient --- otherwise known as a shock. Conservation of mass, energy and momentum across the jump (the Rankine-Hugoniot conditions) reveals a wonderful quirk of physics --- that differential equations with apparently no dissipation can result in irreversible conversion of kinetic energy into heat (i.e., entropy production).
 
  All numerical methods for solving the Euler equations must therefore deal with the inevitable formation of discontinuities. That is, they must provide a mechanism for irreversible dissipation at shock fronts. SPH is no exception.
  
\section{In the beginning, there was artificial viscosity}
   The simplest way to do so was proposed by Von Neumann \& Richtmyer in 1950 \cite{vonneumannrichtmyer50} --- add an `artificial viscosity' term that acts on the shortest resolvable length scale in the calculation, performing the role that microscopic viscosity plays in Nature. In their model the viscosity is artificial because the coefficient depends on resolution. They wrote this as a modification to the pressure $P' = P + q$, where
\begin{equation}
q  = -\frac{(c \Delta x)^2}{V} \frac{\partial U}{\partial x} \left\vert \frac{\partial U}{\partial x}  \right\vert,
\end{equation}
where $U$ is the velocity, $V$ is the specific volume, $\Delta x$ is the grid spacing and $c$ is a dimensionless coefficient of order unity.

 The main objection to the Von-Neumann Richtmyer approach is that the non-linear term, by itself, is not sufficient to completely eliminate post-shock oscillations (at least, not unless large values of $c$ are used, e.g. \cite{rss13}). To eliminate these (as we will demonstrate) requires adding a term that is linear in the viscosity gradient \cite{landshoff55,wilkins80}. The SPH formulation of artificial viscosity, first formulated by Monaghan \& Gingold \cite{monaghangingold83} is based on these principles, adding an additional term to the SPH equation of motion in the form \cite{monaghan85,monaghan92}
\begin{equation}
\left ( \frac{{\rm d}\bm{v}_a}{{\rm d}t} \right)_{\rm AV} = -\sum_b m_b  \frac{-\alpha \overline{c}_{ab} \mu_{ab} + \beta \mu_{ab}^2}{\overline{\rho}_{ab}} \nabla_a W_{ab}, \label{eq:avold}
\end{equation}
where 
\begin{equation}
\mu_{ab} \equiv \begin{cases}
\frac{\bm{v}_{ab}\cdot \bm{r}_{ab}}{r_{ab}^2 + \epsilon h^2} & \bm{v}_{ab}\cdot \bm{r}_{ab} < 0; \\
0 & \bm{v}_{ab}\cdot \bm{r}_{ab} \geq 0,
\end{cases}
\end{equation}
and $\bm{v}_{ab} \equiv \bm{v}_a - \bm{v}_b$, the overbar denotes an average e.g. $\overline{c}_{ab} \equiv \frac12(c_a + c_b)$, and $c$ is the sound speed. This provides both the non-linear Von Neumann-Richtmyer term (with coefficient $\beta$) and a term linear in the velocity gradient and resolution length $h$ (with coefficient $\alpha$). Taking the velocity difference only along the line of sight ensures conservation of both linear and angular momentum in SPH, while the corresponding heating term in the SPH thermal energy equation guarantees a positive definite contribution to the entropy, providing the necessary irreversible dissipation.

\section{Let there be Godunov}
 The requirement for a linear viscosity term is a simple consequence of Godunov's theorem: \emph{Linear numerical schemes for solving partial differential equations that do not generate new extrema (i.e., preserve monotonicity), can be at most first-order accurate} \cite{godunov59}. First order accuracy at shocks is thus a necessary condition for eliminating post-shock oscillations. The problem is that adding a linear viscosity term everywhere (our $\alpha$ term in SPH) ruins the convergence of the entire method even where no shocks or discontinuities are present (i.e., in \emph{smooth flow}).

 The standard approach to grid-based computational fluid dynamics starts with the Euler equations in conservation form
\begin{equation}
\frac{\partial{\bm{u}}}{\partial t} + \nabla \cdot \left[ \bm{F} (\bm{u}) \right] = 0.
\end{equation}
Discretising this using a finite volume approach, e.g. in one dimension with a first order discretisation in time gives
\begin{equation}
\frac{\bm{u}^{n+1}_i - \bm{u}^n_i}{\Delta t} =  -\left[ \frac{\bm{F}^*_{i+\frac12} - \bm{F}^*_{i-\frac12}} {\Delta x} \right],
\end{equation}
where $i$ is the spatial cell index and $n$ is the timestep number. The key insight is to compute the flux at the \emph{interface} between two adjacent cells (in this case $i$ and $i+1$ or $i$ and $i-1$). The idea of Godunov methods is to compute this flux by considering adjoining cells as the left and right states of a Riemann problem.

 In Godunov's original method, one considers every pair of cells to be a shock. This results in a first order scheme by virtue of Godunov's theorem. The way around this \cite{van-leer79,colellawoodward84} is to no longer consider the values $\bm{u}$ in the left and right states as constant, but instead to use the gradients to \emph{reconstruct} the values of $\bm{u}$ at the interface. In smooth flow, the reconstruction is exact and no dissipation occurs --- because there is no jump in the fluid variables to produce irreversible dissipation. Reconstruction thus restores higher order convergence of the scheme in smooth flow. However, by Godunov's theorem it is impossible to provide second order accuracy at shocks without re-introducing post-shock oscillations. So reconstruction is done with `slope-limited' gradients to ensure that the scheme remains monotone (i.e., does not introduce new maxima or minima). Slope limiters thus revert to the required first order accuracy when the gradients are discontinuous.

 When reconstruction is employed, it is no longer necessary to \emph{exactly} solve the Riemann problem. For many applications an approximate solution will suffice. The simplest `approximate Riemann solver' uses the `local Lax-Friedrichs' flux given by
\begin{equation}
\bm{F}^* = \frac12 \left[ \bm{F}(\bm{u}_L) + \bm{F}(\bm{u}_R) \right] - \frac{v_{\rm sig}}{2} (\bm{u}_R - \bm{u}_L), \label{eq:llf}
\end{equation}
where $v_{\rm sig}$ is the maximum signal speed, corresponding to the maximum eigenvalue of the Jacobian flux matrix. Approximate Riemann solvers are more dissipative than exact solvers because, for example, the above assumes that every discontinuity is a shock, rather than considering shocks, rarefactions and contact discontinuities separately. The advantage is that one merely needs to know $v_{\rm sig}$ to generalise the method to any set of hyperbolic PDEs, which is trivial compared to finding an exact solution to the Riemann problem in every case.

\section{And there was Godunov (sort of)}
 Monaghan \cite{monaghan97} used the principles above to generalise the SPH artificial viscosity. Considering SPH equations in Lagrangian conservative form
\begin{equation}
\frac{{\rm d}{\bm{u}}}{{\rm d}t} = - \frac{\nabla \cdot \bm{F} (\bm{u})}{\rho},
\end{equation}
we can write corresponding SPH equations in the form corresponding to (\ref{eq:llf}) by considering each pair of particles as the left and right states. We obtain
\begin{eqnarray}
\frac{{\rm d}{\bm{u}_a}}{{\rm d}t} & = & - \sum_b m_b \left[ \frac{\bm{F}_a}{\Omega_a \rho_a^2} \cdot \nabla_a W_{ab}(h_a) +  \frac{\bm{F}_b}{\Omega_a \rho_a^2} \cdot \nabla_a W_{ab}(h_b) \right] \nonumber \\
&  & - \sum_b \frac{m_b}{\overline{\rho}_{ab}} \overline{v}_{\rm sig} (\bm{u}_a - \bm{u}_b) \hat{\bm{r}}_{ab} \cdot \overline{\nabla_a W}_{ab} ,
\end{eqnarray}
where the bar denotes an average. For hydrodynamics we have
\begin{equation}
\bm{u} = \left[ \begin{array}{c} \bm{v} \\ e \end{array} \right]; \hspace{1cm} \bm{F} = \left[ \begin{array}{c} P \bm{I} \\ P\bm{v} \end{array} \right],
\end{equation}
where $\bm{v}$ is the velocity, $e = \frac12 v^2 + u$ is the specific total energy, $P$ is the pressure and $u$ is the specific thermal energy.

\subsection{Artificial viscosity}
To conserve angular momentum it is necessary to only dissipate the velocity component along the line of sight (but see \cite{pricemonaghan05} who experimented with a version using the total velocity jump), giving equations of motion with artificial viscosity term in the form
\begin{eqnarray}
\frac{{\rm d}{\bm{v}_a}}{{\rm d}t} & = & - \sum_b m_b \left[ \frac{P_a}{\Omega_a \rho_a^2} \nabla_a W_{ab}(h_a) +  \frac{P_b}{\Omega_a \rho_a^2} \nabla_a W_{ab}(h_b) \right] \nonumber \\
&  & - \sum_b \frac{m_b}{\overline{\rho}_{ab}} \overline{v}_{\rm sig} (\bm{v}_a - \bm{v}_b) \cdot \hat{\bm{r}}_{ab} \overline{\nabla_a W}_{ab}.
\end{eqnarray}
For hydrodynamics the signal speed can be expressed as
\begin{equation}
\overline{v}_{\rm sig} = \alpha \frac{c_{a} + c_{b}}{2} - \beta \bm{v}_{ab} \cdot \hat{\bm{r}}_{ab},
\end{equation}
thus providing the same linear and quadratic viscosity terms as in (\ref{eq:avold}). Artificial viscosity may thus be viewed as an approximate Riemann solver.

\subsection{Artificial thermal conductivity}
The other implication is in the energy equation, which in the above formulation corresponds to \cite{chowmonaghan97,monaghan97}
\begin{eqnarray}
\frac{{\rm d}{e_a}}{{\rm d}t} & = & - \sum_b m_b \left[ \frac{P_a \bm{v}_b}{\Omega_a \rho_a^2} \cdot \nabla_a W_{ab}(h_a) +  \frac{P_b \bm{v}_a}{\Omega_a \rho_a^2} \cdot \nabla_a W_{ab}(h_b) \right] \nonumber \\
&  & - \sum_b \frac{m_b}{\overline{\rho}_{ab}} \overline{v}_{\rm sig} (e^*_a - e^*_b) \hat{\bm{r}}_{ab} \cdot \overline{\nabla_a W}_{ab},
\end{eqnarray}
where $e^*_a = \frac12 (\bm{v}_a \cdot \hat{\bm{r}}_{ab}) + u_a$ is the energy considering the kinetic energy only along the line of sight joining the particles. The kinetic energy jump provides the usual viscous heating term in the thermal energy equation, but the thermal energy jump $(u_a - u_b)$ provides an artificial thermal conductivity which turns out to be necessary at contact discontinuities (see \cite{price08}). The corresponding thermal energy equation (it is unnecessary to evolve the total energy in practice) is given by
\begin{eqnarray}
\frac{{\rm d}{u_a}}{{\rm d}t}  & = & \frac{{\rm d}{e_a}}{{\rm d}t} - \bm{v}_a \cdot \frac{{\rm d}{\bm{v}_a}}{{\rm d}t},  \nonumber \\
& = & \frac{P_a}{\Omega_a \rho_a} \sum_b m_b \bm{v}_{ab} \cdot \nabla_a W_{ab}(h_a)  \label{eq:dudt} \\
& - & \sum_b \frac{m_b}{\overline{\rho}_{ab}} \overline{v}_{\rm sig} \left[ \frac12 (\bm{v}_{ab}\cdot \bm{r}_{ab})^2 + (u_b - u_a) \right] \overline{F}_{ab},\nonumber 
\end{eqnarray}
where $\overline{F}_{ab} \equiv \hat{\bm{r}}_{ab} \cdot \overline{\nabla_a W}_{ab}$. The corresponding contribution to the entropy is
\begin{equation}
T_a \frac{{\rm d}s_a}{{\rm d}t} = \frac{{\rm d}{u_a}}{{\rm d}t} - \frac{P_a}{\rho_a^2} \frac{{\rm d}{\rho_a}}{{\rm d}t},
\end{equation}
which equals the last term in (\ref{eq:dudt}) which is always positive for viscosity. The proof that the conductivity term gives positive definite entropy is given in the Appendix of \cite{pricemonaghan04a}.

\begin{figure}
\centering
\includegraphics[width=\columnwidth]{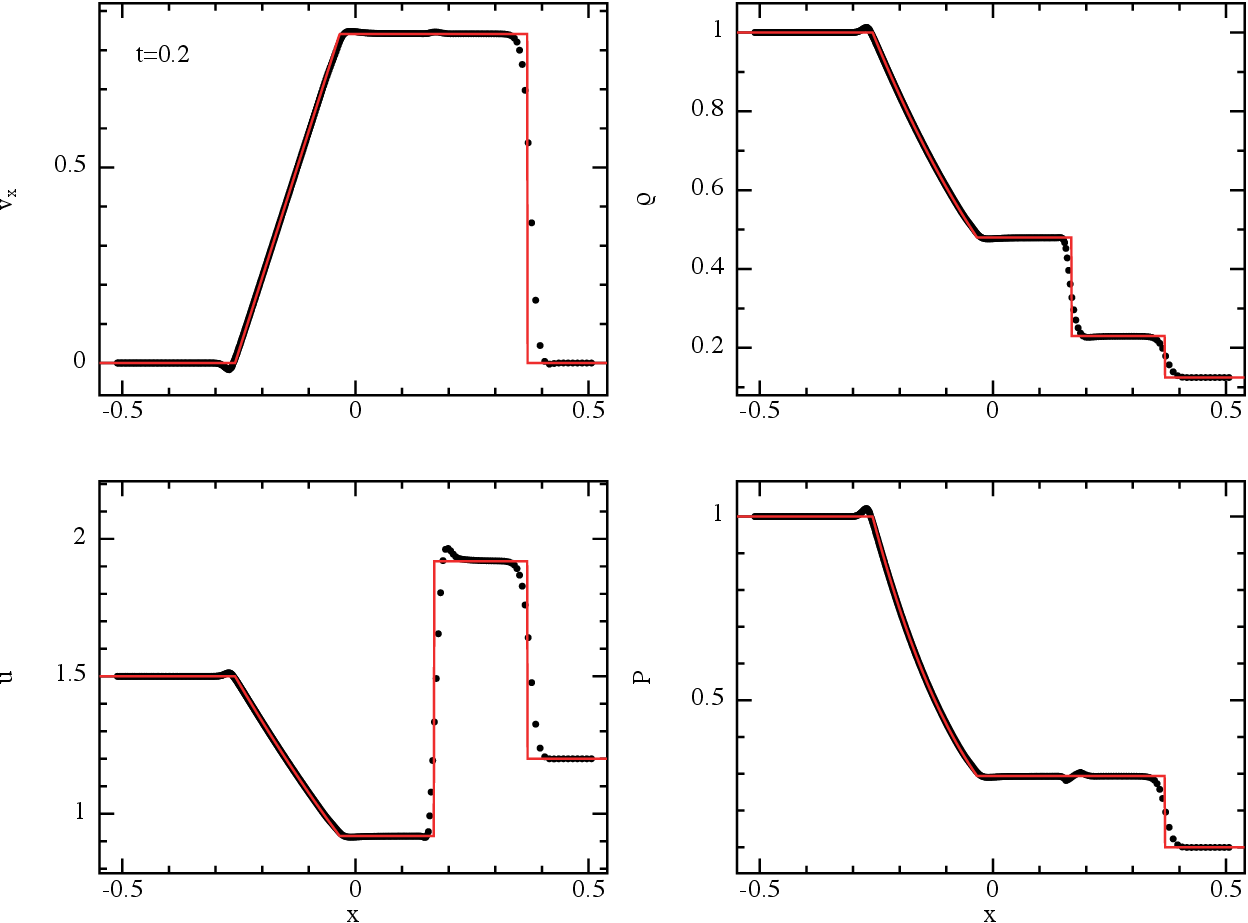}
% where an .eps filename suffix will be assumed under latex,
% and a .pdf suffix will be assumed for pdflatex
\caption{1D Sod shock tube computed using artificial viscosity with constant coefficients $\alpha = 1$ and $\beta = 2$.}
\label{fig:sodstd}
\end{figure}

\begin{figure}
\centering
\includegraphics[width=\columnwidth]{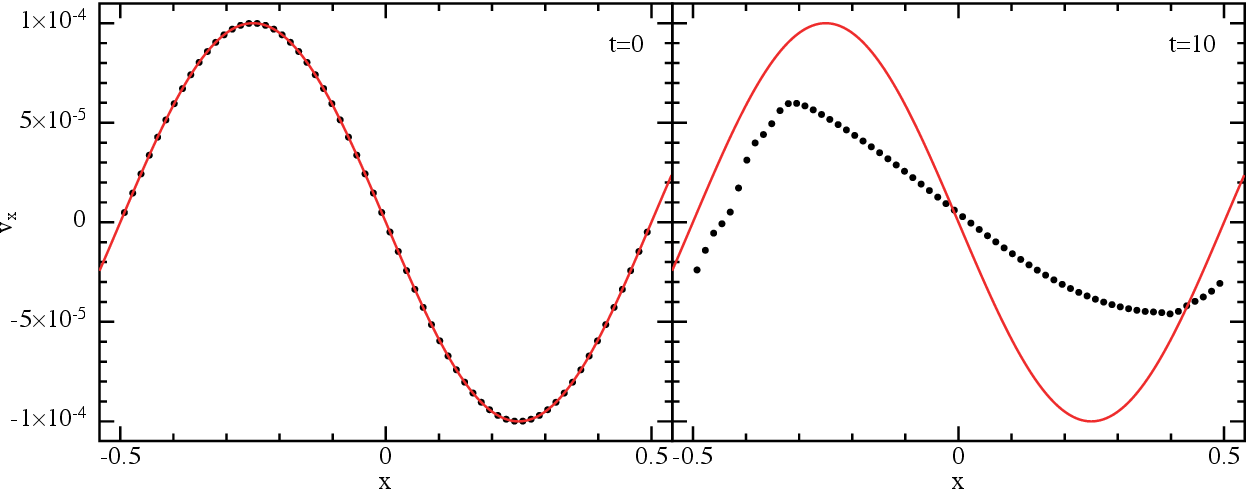}
% where an .eps filename suffix will be assumed under latex,
% and a .pdf suffix will be assumed for pdflatex
\caption{Linear wave in 1D computed using artificial viscosity with constant coefficients $\alpha = 1$ and $\beta = 2$. We employ 64 particles in 1D.}
\label{fig:wavestd}
\end{figure}

\begin{figure}[t]
\centering
\includegraphics[width=\columnwidth]{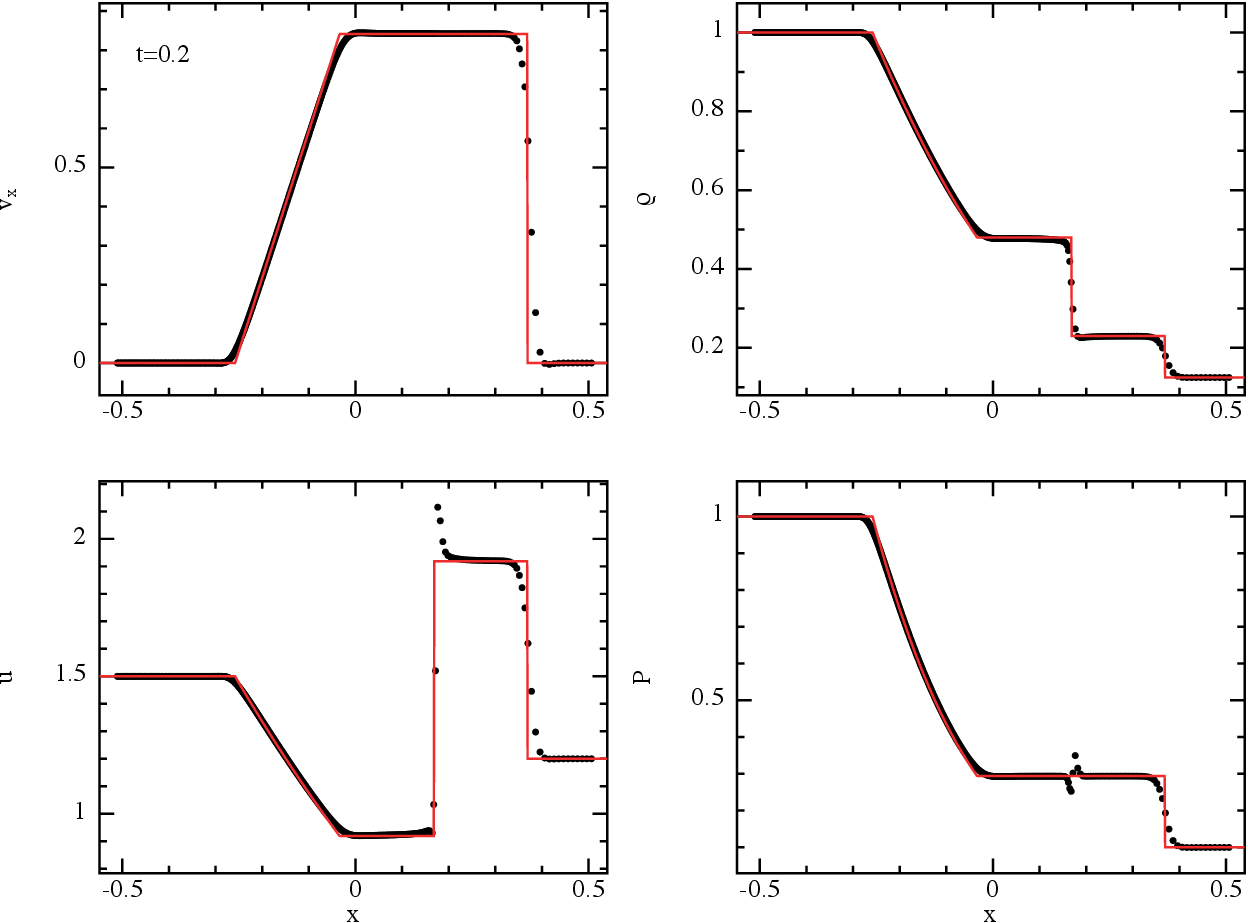}
% where an .eps filename suffix will be assumed under latex,
% and a .pdf suffix will be assumed for pdflatex
\caption{Sod shock tube computed with Godunov SPH assuming piecewise constant reconstruction (Eq.~\ref{eq:pstar}).}
\label{fig:sodgsph}
\end{figure}

\begin{figure}
\centering
\includegraphics[width=\columnwidth]{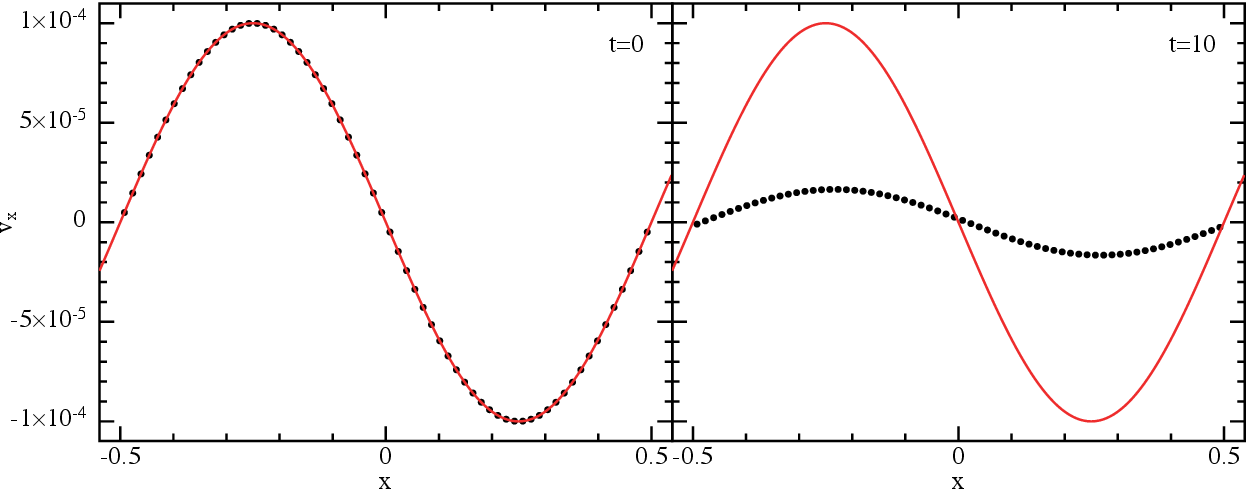}
% where an .eps filename suffix will be assumed under latex,
% and a .pdf suffix will be assumed for pdflatex
\caption{Linear wave computed with Godunov SPH assuming piecewise constant reconstruction (Eq.~\ref{eq:pstar}). We employ 64 particles in 1D.}
\label{fig:wavegsph}
\end{figure}

\subsection{Some minor tweaks}
In practice \cite{priceetal18a} we write the momentum and thermal energy contributions as a $P + q$ term, i.e.
\begin{equation}
\frac{{\rm d}{\bm{v}_a}}{{\rm d}t}  = - \sum_b m_b \left[ \frac{P_a + q_a}{\Omega_a \rho_a^2} \nabla_a W_{ab}(h_a) +  \frac{P_b+q_b}{\Omega_a \rho_a^2}\nabla_a W_{ab}(h_b) \right],
\end{equation}
where
\begin{equation}
q_a = \begin{cases}
 -\frac12 \alpha_a \rho_a v_{{\rm sig}, a} ( \bm{v}_{ab} \cdot \hat{\bm{r}}_{ab}) &  ( \bm{v}_{ab} \cdot \hat{\bm{r}}_{ab})  < 0; \\
 0 &  \text{otherwise}
 \end{cases}
\end{equation}
This is mainly to avoid the need for arbitrary averages (the choice of averaging has no obvious effect on the results). We also use a separate signal speed for conductivity \cite{price08}, namely $v_{\rm sig}^u = \sqrt{ {\vert P_a - P_b \vert} / {\overline{\rho}_{ab}}}$, which is designed to eliminate the pressure `blip' at the contact discontinuity. This also results in second order artificial conductivity (since $v_{\rm sig}^u \propto h$). This means we can simply use a constant coefficient $\alpha_u = 1$ while retaining second order convergence.

\subsection{Some shocking results}
 Figure~\ref{fig:sodstd} shows the results of the standard Sod shock tube in SPH (the setup is as described in \cite{price08}, using 450 equal mass particles in 1D) computed using the viscosity and conductivity terms above with $\alpha = \alpha_u = 1$ and $\beta = 2$. No post-shock oscillations are evident. Both the shock and contact discontinuity are spread over several smoothing lengths [for all tests we use the $M_6$ quintic kernel with radius $3h$ and $h = 1.0 (m/\rho)$].
 
  Figure~\ref{fig:wavestd} shows a linear sound wave test with the same parameters, using 64 particles in 1D. Right panel shows solution after 10 wave periods. We observe that the amplitude of the wave is reduced by almost half, while the shape is distorted because we apply viscosity only where $\nabla\cdot \bm{v} < 0$.

\section{Godunov SPH}
 Other authors have attempted a more exact approach to a Godunov formulation of SPH. Inutsuka \cite{inutsuka02} proposed to restore integral conservation laws by considering the equations of motion convolved with the SPH kernel, i.e.
\begin{equation}
\int \left ( \frac{{\rm d}\bm{v}}{{\rm d}t} \right) W(\bm{r} - \bm{r}') {\rm d}V' = - \int \left ( \frac{\nabla P}{\rho} \right) W(\bm{r} - \bm{r}') {\rm d}V',
\end{equation}
deriving semi-discrete SPH equations of motion given by
\begin{equation}
\frac{{\rm d}\bm{v}_a}{{\rm d}t} = - \sum_b m_b \int \frac{P(\bm{r})}{\rho^2(\bm{r})} \left[ \nabla_a - \nabla_b \right] W(\bm{r} - \bm{r}_a) W(\bm{r} - \bm{r}_b) {\rm d}V'.
\end{equation}
The Riemann solver arrives by replacing the momentum flux (pressure) by $P^*$, the flux between the two particles obtained as the solution to the Riemann problem, giving
\begin{equation}
\frac{{\rm d}\bm{v}_a}{{\rm d}t} = - \sum_b m_b \int \frac{P^*_{a,b}}{\rho^2(\bm{r})} \left[ \nabla_a - \nabla_b \right] W(\bm{r} - \bm{r}_a) W(\bm{r} - \bm{r}_b) {\rm d}V'. \label{eq:inutsuka}
\end{equation}
 The algorithm is complicated by the need to further discretise the remaining integral on the right hand side, which is non-trivial. The final aspect is to restore higher order convergence away from shocks by `reconstructing' the left and right states of the Riemann problem. That is, using the gradients of $\rho$, $\bm{v}$ and $P$ evaluated at each particle to interpolate to the midpoint. Monotonicity constraints can be enforced by setting the velocity gradients to zero if they have opposing signs.
 
 \begin{figure}[t]
\centering
\includegraphics[width=\columnwidth]{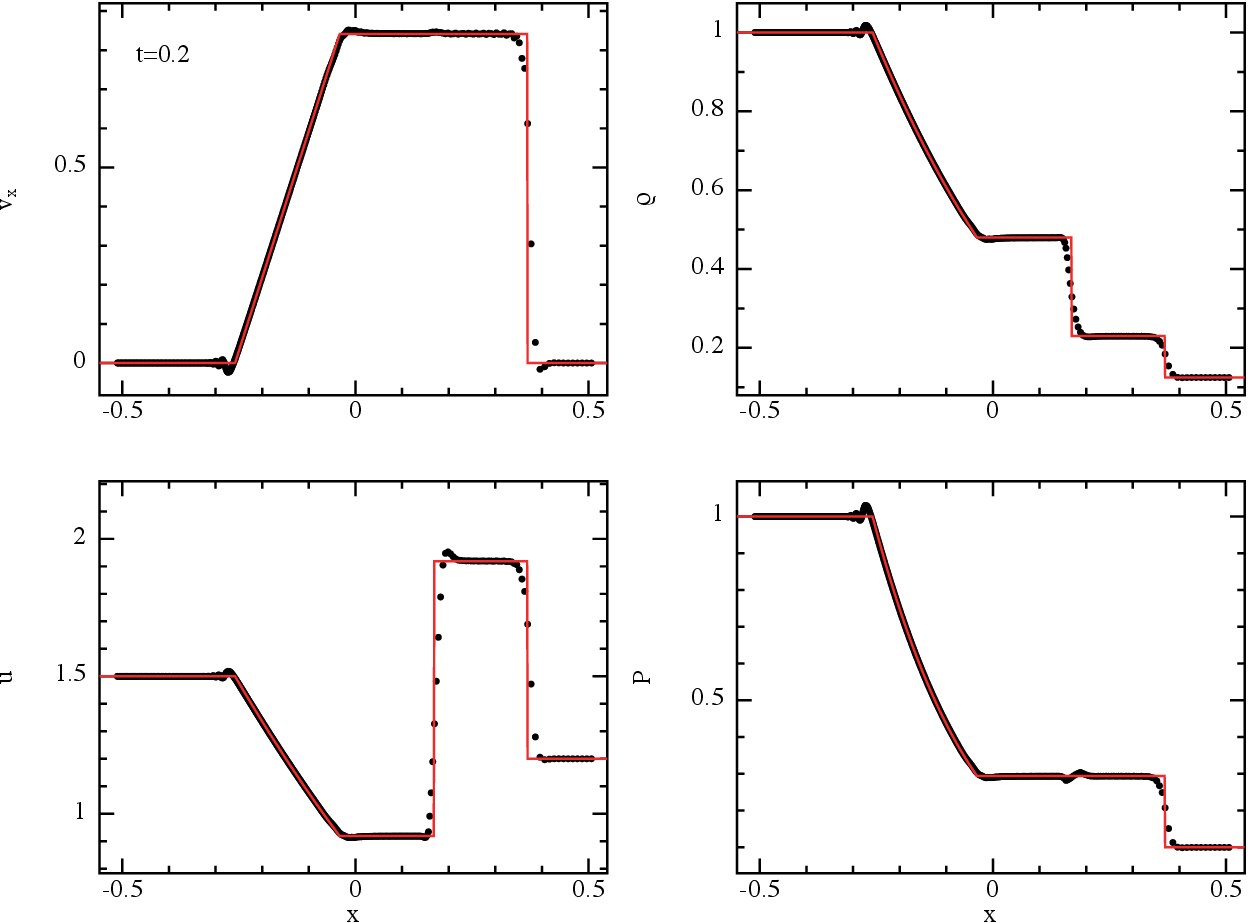}
% where an .eps filename suffix will be assumed under latex,
% and a .pdf suffix will be assumed for pdflatex
\caption{1D Sod shock tube computed using artificial viscosity with $\alpha \in [0.1,1]$ and $\beta = 2$ using the Morris \& Monaghan (1997) switch \cite{morrismonaghan97}.}
\label{fig:sodavlim}
\end{figure}

\begin{figure}
\centering
\includegraphics[width=\columnwidth]{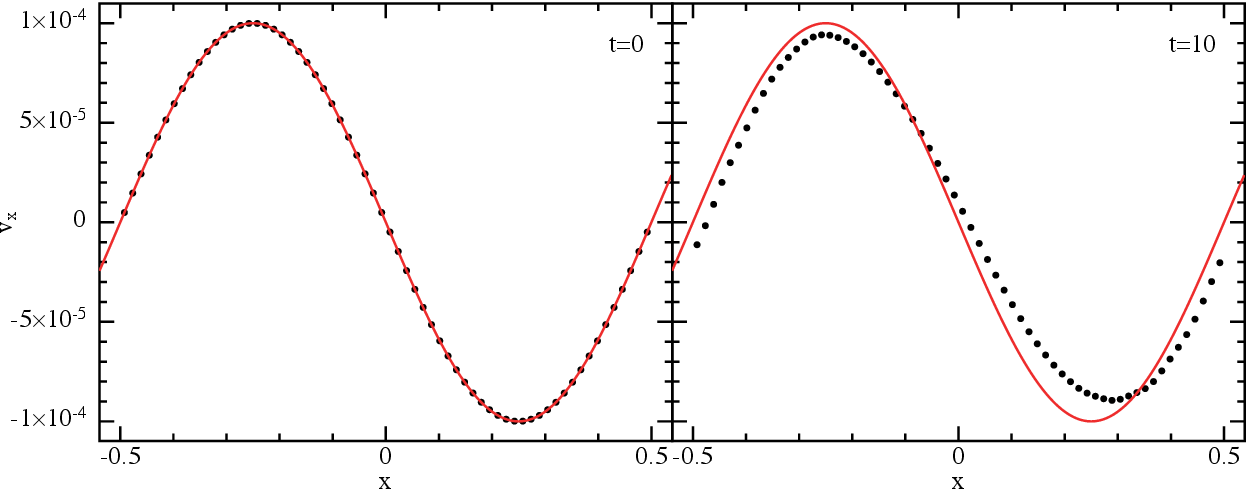}
% where an .eps filename suffix will be assumed under latex,
% and a .pdf suffix will be assumed for pdflatex
\caption{Linear sound wave in 1D computed using artificial viscosity with $\alpha \in [0.1,1]$ and $\beta = 2$ using the Morris \& Monaghan (1997) switch \cite{morrismonaghan97}.}
\label{fig:waveavlim}
\end{figure}

\begin{figure}[t]
\centering
\includegraphics[width=\columnwidth]{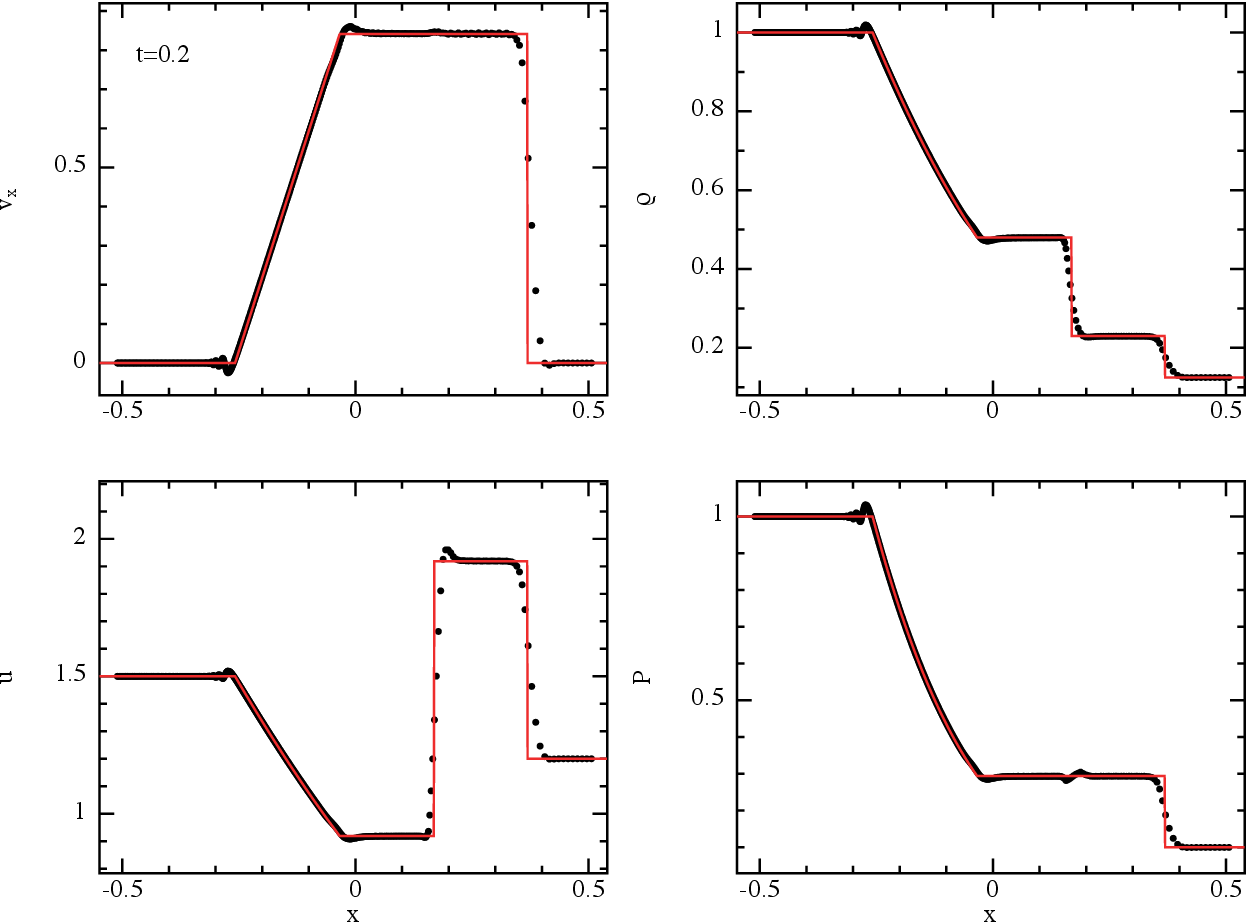}
% where an .eps filename suffix will be assumed under latex,
% and a .pdf suffix will be assumed for pdflatex
\caption{1D Sod shock tube computed using artificial viscosity with $\alpha = 0$ and $\beta = 10$.}
\label{fig:sodbeta10}
\end{figure}

\begin{figure}
\centering
\includegraphics[width=\columnwidth]{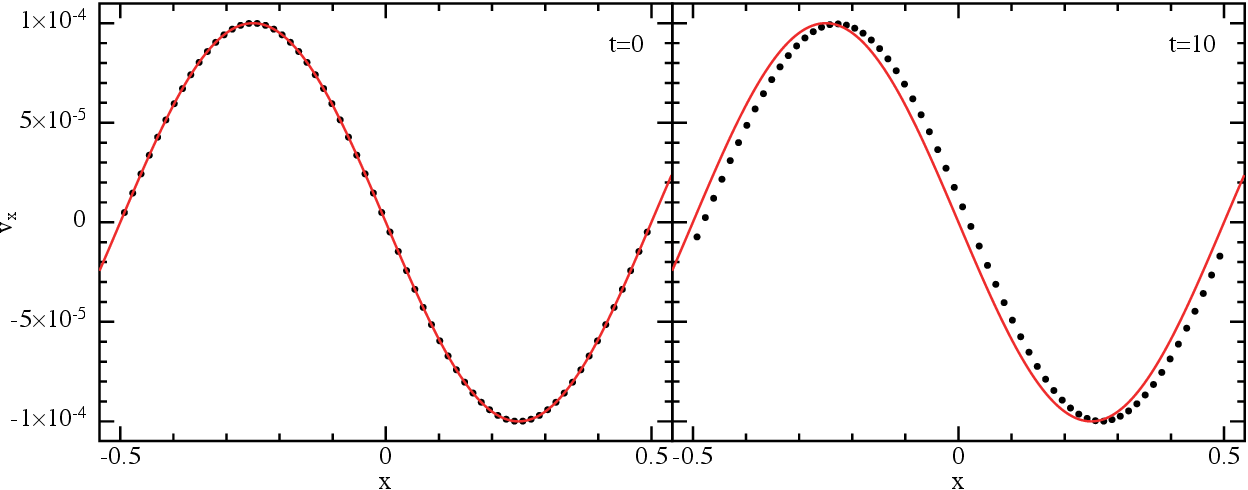}
% where an .eps filename suffix will be assumed under latex,
% and a .pdf suffix will be assumed for pdflatex
\caption{Linear sound wave in 1D computed using artificial viscosity with $\alpha = 0$ and $\beta = 10$. Look mum, no switches!}
\label{fig:wavebeta10}
\end{figure}

 Cha \& Whitworth \cite{chawhitworth03} demonstrated that the `Riemann solver' and `complicated second integral' parts of the algorithm can be separated. For example, one can simply use
\begin{equation}
\frac{{\rm d}\bm{v}_a}{{\rm d}t} = - \sum_b m_b \left[ \frac{P^*_{a,b}}{\Omega_a \rho_a^2}  \nabla_a W_{ab}(h_a) + \frac{P^*_{a,b}}{\Omega_b \rho_b^2}  \nabla_a W_{ab}(h_b) \right]. \label{eq:pstar}
\end{equation}
Figure~\ref{fig:sodgsph} shows the Sod shock tube computed with Godunov SPH according to (\ref{eq:pstar}). We see that even this simple scheme successfully removes post-shock oscillations. The problem is that without implementing reconstruction --- which in \cite{inutsuka02} and \cite{chawhitworth03} is tied up with evaluating the complicated `second integral' --- this scheme remains first order away from shocks.  Figure~\ref{fig:wavegsph} shows the corresponding 1D linear wave solution, but we are left with significant damping of small amplitude waves, worse than for our standard artificial viscosity scheme (comparing Fig.~\ref{fig:wavegsph} to Fig.~\ref{fig:wavestd}, see also \cite{price08}).
 
 A second issue is the use of an exact Riemann solver. Approximate solvers may suffice --- and are necessary if more complicated physics is employed, e.g. magnetic fields.

\begin{figure}[t]
\centering
\includegraphics[width=\columnwidth]{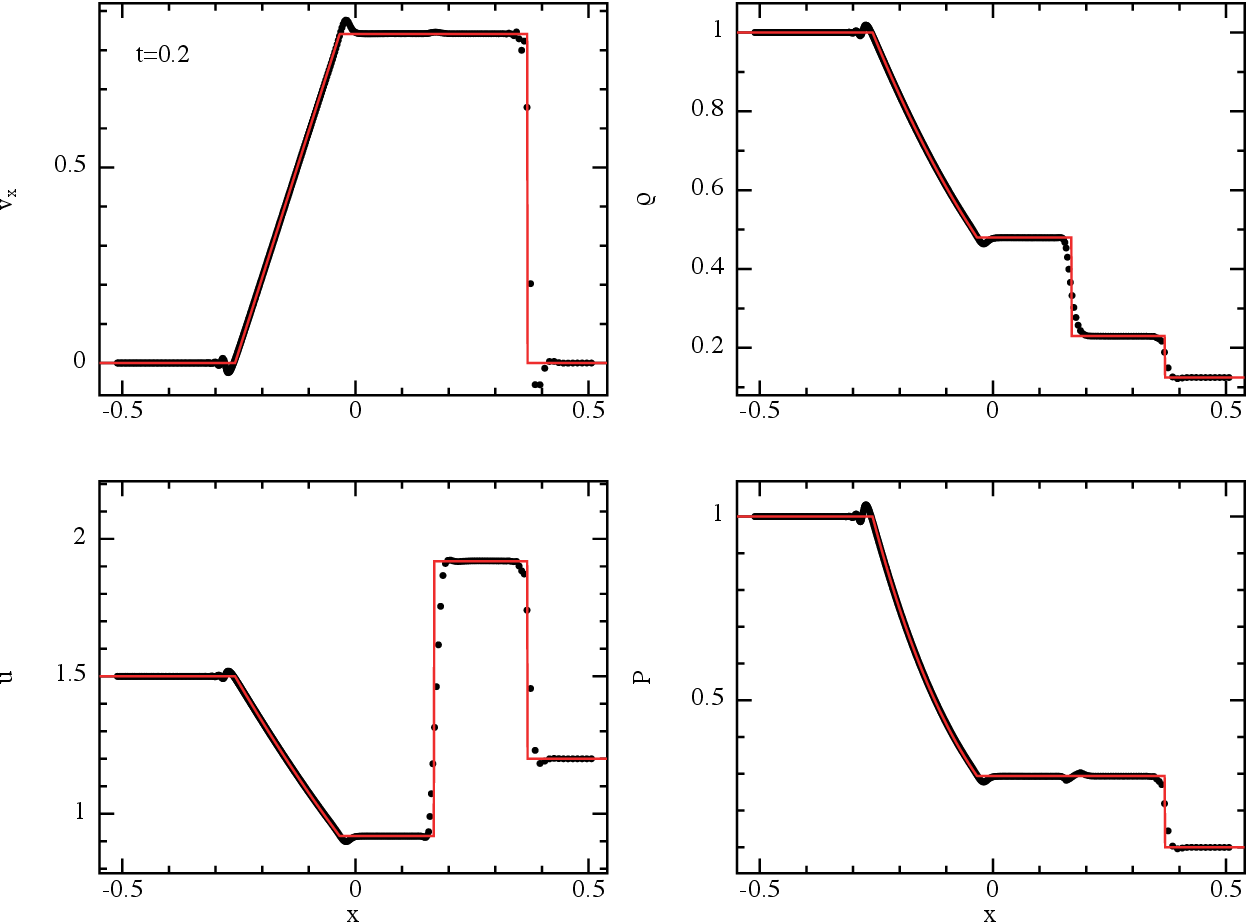}
% where an .eps filename suffix will be assumed under latex,
% and a .pdf suffix will be assumed for pdflatex
\caption{1D Sod shock tube computed using artificial viscosity with $\alpha = 1$ and $\beta = 2$ applying reconstruction with no slope limiter.}
\label{fig:sodrecon0}
\end{figure}

\begin{figure}
\centering
\includegraphics[width=\columnwidth]{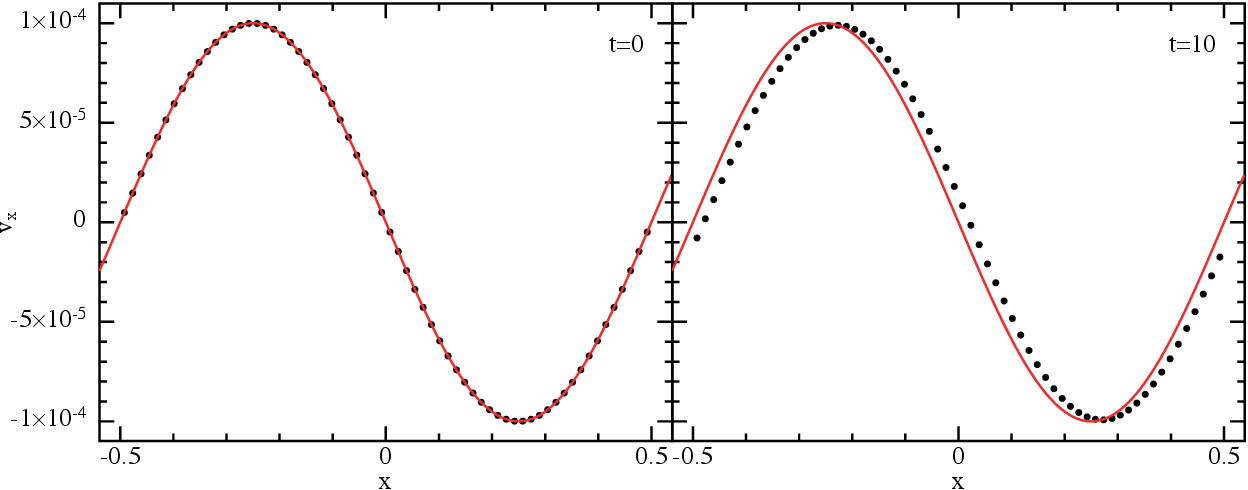}
% where an .eps filename suffix will be assumed under latex,
% and a .pdf suffix will be assumed for pdflatex
\caption{Linear sound wave in 1D computed using artificial viscosity with $\alpha = 1$ and $\beta = 2$ applying reconstruction with no slope limiter.}
\label{fig:waverecon0}
\end{figure}

\begin{figure}[t]
\centering
\includegraphics[width=\columnwidth]{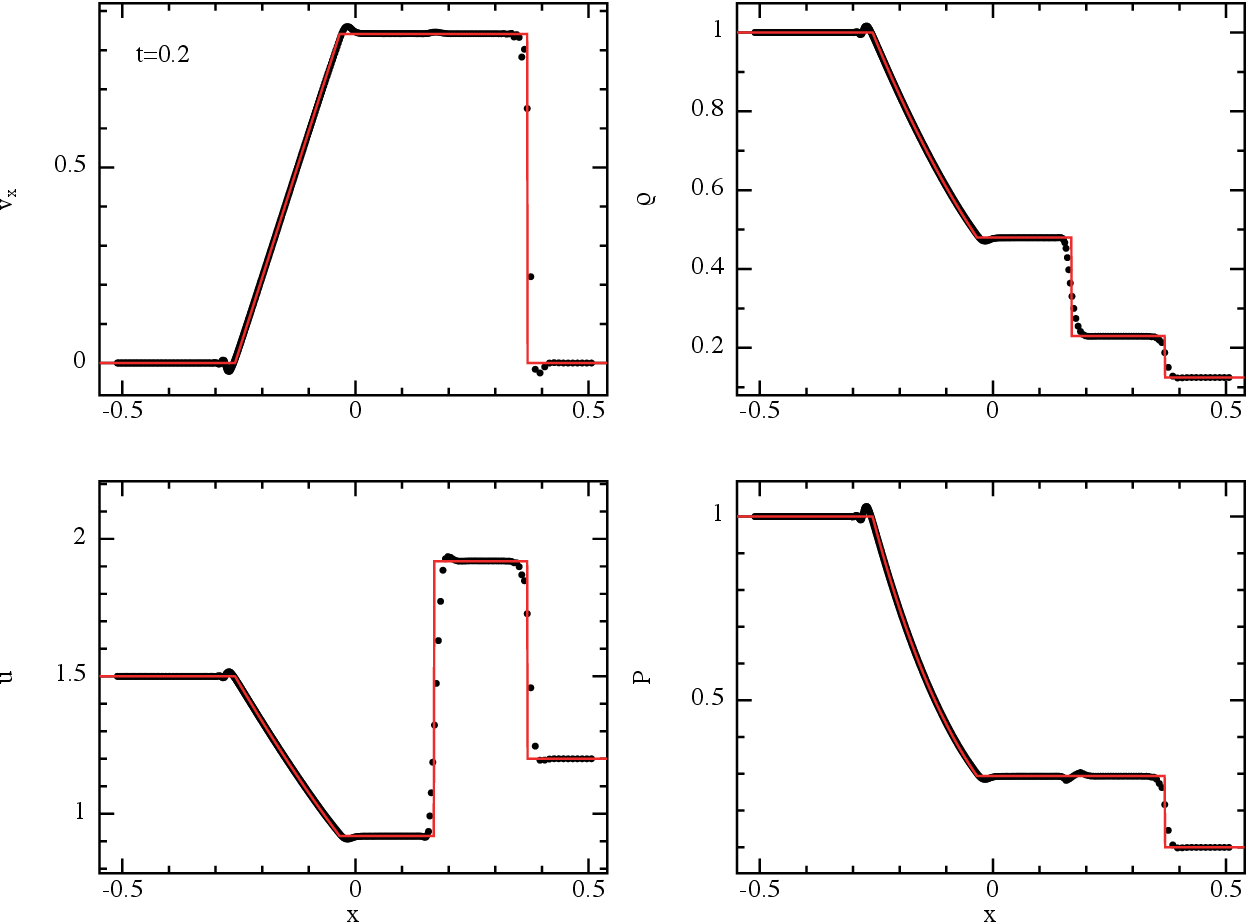}
% where an .eps filename suffix will be assumed under latex,
% and a .pdf suffix will be assumed for pdflatex
\caption{1D Sod shock tube computed using artificial viscosity with $\alpha = 1$ and $\beta = 2$ applying reconstruction with slope limiter 2.}
\label{fig:sodrecon2}
\end{figure}

\begin{figure}
\centering
\includegraphics[width=\columnwidth]{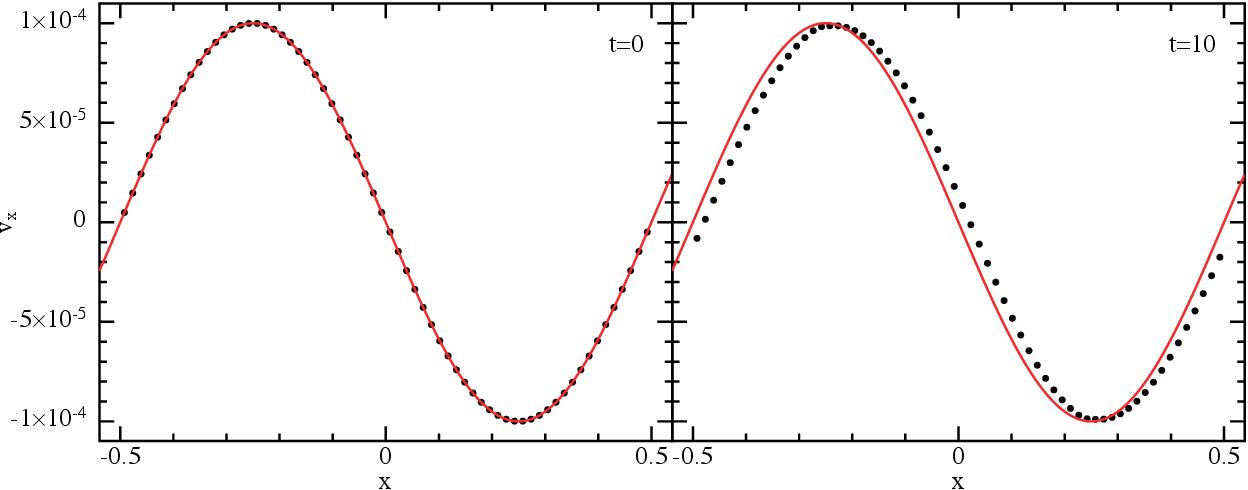}
% where an .eps filename suffix will be assumed under latex,
% and a .pdf suffix will be assumed for pdflatex
\caption{Linear sound wave in 1D computed using artificial viscosity with $\alpha = 1$ and $\beta = 2$ applying reconstruction with slope limiter 2.}
\label{fig:waverecon2}
\end{figure}

\section{Let there be switches...}
 In both of the above cases, the medicine works (no post-shock oscillations), but there are side effects (excess dissipation away from shocks). The challenge is to preserve shock capturing while reducing dissipation away from shocks. Morris \& Monaghan \cite{morrismonaghan97} proposed evolving $\alpha$ for each particle using
\begin{equation}
\frac{{\rm d}\alpha_a}{{\rm d}t} = \mathcal{S} - \frac{\alpha -\alpha_{\rm min}}{\tau}, \label{eq:alpha}
\end{equation}
where $\mathcal{S} = \max[0, -(\nabla\cdot\bm{v})_a]$ is the shock detector and $\tau \equiv 0.1 h_a / v_{\rm sig}$ is the decay timescale, damping $\alpha$ to $\alpha_{\min}$ over $\sim 5$--10 smoothing lengths.

 Figure~\ref{fig:sodavlim} shows the Sod shock with individual $\alpha$'s evolved using (\ref{eq:alpha}) with $\alpha_{\min} = 0.1$. The shock solution remains similar, but damping of the linear wave (Figure~\ref{fig:waveavlim}) is greatly reduced. This provides an effective solution, demonstrated by \cite{puriramachandran14} to be superior to numerous other proposed methods. 
 
 Cullen \& Dehnen pointed out several flaws with the Morris \& Monaghan switch. The first is that $\alpha$ rises too slowly in (\ref{eq:alpha}) at strong shocks, leading to some visible post-shock oscillations. The second is that the viscosity remains switched on for adiabatic compression and expansion, where $\nabla\cdot\bm{v} \neq 0$ but no shock is present. They therefore proposed a more complicated switch, using the time derivative of the velocity divergence as the shock detector. They define
 \begin{equation}
A_{a} = \xi_{a} \max \left[-\frac{{\rm d}}{{\rm d}t}(\nabla\cdot{\bm v}_{a}), 0 \right] ,
\label{eq:avsource}
\end{equation}
where $\xi_{a}$ is the ratio of the velocity divergence to the sum of the divergence and curl (akin to the Balsara switch, \cite{balsara95}) and
 \begin{equation}
\frac{{\rm d}}{{\rm d}t}\left( \nabla\cdot \bm{v}_a \right) = (\nabla\cdot\bm{a})_a - \frac{\partial v_{a}^{i}}{\partial x_{a}^{j}} \frac{\partial v_{a}^{j}}{\partial x_{a}^{i}}.
\label{eq:ddivvdt}
\end{equation}
In \cite{priceetal18a} we used a slightly modified version of their scheme, setting $\alpha$ according to
\begin{equation}
\alpha_{{\rm loc}, a} =\min\left( \frac{10 h_{a}^{2} A_{a}}{c_{{\rm s}, a}^{2}}, \alpha_{\rm max} \right), \label{eq:alphaloc}
\end{equation}
finally setting $\alpha$ to the maximum of $\alpha_{\rm loc}$ and the $\alpha$ evolved according to
\begin{equation}
\frac{{\rm d}\alpha_{a}}{{\rm d}t} = - \frac{(\alpha_{a} - \alpha_{{\rm loc}, a})}{\tau_{a}},
\end{equation}
where $\tau$ is the decay timescale as previously. 
 
 The above switch involves, amongst other things, computing the divergence of the acceleration, which requires sending the previous acceleration into the density routine. This gets ugly in the code. Worse is that we essentially had to hack (\ref{eq:alphaloc}) to give acceptable results in 3D, with the uneasy feeling that our parameter settings are empirical and arbitrary.

%After the text edit has been completed, the paper is ready for the template. Duplicate the template file by using the Save as command, and use the naming convention prescribed by the workshop for the name of your paper: ``SPHERICIV\_LastName1stAuthor\_LastName2ndAuthor.pdf''.

% Reminder: the "draftcls" or "draftclsnofoot", not "draft", class option
% should be used if it is desired that the figures are to be displayed while
% in draft mode.

% An example of a floating figure using the graphicx package.
% Note that \label must occur AFTER (or within) \caption.
% For figures, \caption should occur after the \includegraphics.
%

% An example of a double column floating figure using two subfigures.
%(The subfigure.sty package must be loaded for this to work.)
% The subfigure \label commands are set within each subfigure command, the
% \label for the overall figure must come after \caption.
% \hfil must be used as a separator to get equal spacing
%
%\begin{figure*}
%%\centerline{\subfigure[Case I]{\includegraphics[width=2.5in]{velocity2}
%% where an .eps filename suffix will be assumed under latex,
%% and a .pdf suffix will be assumed for pdflatex
%\hfil
%%\subfigure[Case II]{\includegraphics[width=2.5in]{velocity3}
%% where an .eps filename suffix will be assumed under latex,
%% and a .pdf suffix will be assumed for pdflatex
%\label{fig_second_case}}}
%\caption{Simulation results}
%\label{fig_sim2}
%\end{figure*}

\section{The road less travelled}
 My uneasiness with the Cullen \& Dehnen switch implemented in {\sc phantom} prompted a rethink. Godunov SPH as it stands has not been widely adopted due to the need to discretise the `second integral' and also the need to solve the Riemann problem exactly, both of which add complexity and computational cost. But if we can interpret the standard artificial viscosity scheme as an approximate Riemann solver anyway, maybe we can take the `best of both worlds'?
 
 What we learn from Finite Volume methods is that switches are not the answer. The road to second order convergence away from shocks lies in reconstructing the `primitive' variables to the interface between cells (particles) so that in smooth flows no `jump' is apparent, and therefore no irreversible dissipation is applied. Godunov's theorem implies that slope limiters are required when reconstructing quantities between particles to ensure that the scheme remains monotonic. In SPH language this means `a positive definite contribution to the entropy'.

\subsection{A brief aside --- second order dissipation without switches}
 From \cite{rss13} I learnt that one can in principle use a pure Von Neumann-Richtmyer viscosity for shock capturing, just with a large coefficient. In SPH this means setting $\alpha = 0$ and using a large $\beta$. This seems crazy-brave. Does it work? 
 
 Fig.~\ref{fig:sodbeta10} shows the Sod shock computed using $\alpha = 0$ and $\beta = 10$, while Fig.~\ref{fig:wavebeta10} shows the linear wave. The short answer is \emph{yes}. We successfully eliminate post-shock oscillations and obtain an essentially undamped sound wave (the remaining phase error arises from the kernel). This also works in 3D but I am not brave enough to proceed with this approach without a more detailed study --- mainly because \cite{landshoff55} introduced linear viscosity to avoid doing exactly this.
 
\subsection{The place where switches go to die}
 I propose a `best of both worlds' approach, where instead of playing with $\alpha$ we simply apply reconstruction to (only) the velocity jump in the viscosity term. That is, we use
\begin{equation}
q^*_a = -\frac12 \alpha_a \rho_a v_{{\rm sig}, a} ( \bm{v}^*_{a} - \bm{v}^*_b) \cdot \hat{\bm{r}}_{ab},
\end{equation}
where $\bm{v}^*$ is the velocity reconstructed to the midpoint of the particle pair $\bm{r}^{*} = \bm{r}_{a} + 0.5 \bm{r}_{ab}$, e.g.
\begin{equation}
\bm{v}^{*}_a = \bm{v}_a + \left(\bm{r}^{*} - \bm{r}_{a}\right)^\beta \frac{\partial \bm{v}_{a}}{\partial \bm{r}_{a}^\beta}.
\end{equation}
We find
\begin{equation}
\bm{v}^{*}_{ab} \cdot \hat{\bm{r}}_{ab} = \bm{v}_{ab} \cdot \hat{\bm{r}}_{ab} + 0.5 \vert r_{ab} \vert  \left( S_{ab} + S_{ba} \right) ,
\end{equation}
where $S_{ab} \equiv \hat{r}_{ab}^\alpha \hat{r}_{ab}^\beta \frac{\partial v_{a}^{\alpha}}{\partial x_{a}^\beta}$. Velocity gradients are computed during the density summation according to
\begin{equation}
\frac{\partial v_a^{\alpha}}{\partial r_a^{\beta}}  = - \frac{1}{\Omega_{a}\rho_{a}} \sum_{b}m_{b} v^{\beta}_{ab}\nabla^{\beta}W_{ab}\left(h_{a} \right).
\end{equation}
Finally, to avoid introducing new maxima or minima, we limit the factor $0.5\left(S_{ab} + S_{ba}\right)$ using a slope limiter, i.e. a function $f\left(S_{ab},S_{ba} \right)$ that prevents the development of numerical oscillations. We use the Van Leer MC limiter, since it provides best compromise between smoothing and dissipation.

 Figs.~\ref{fig:sodrecon0} shows the results of this modification in the Sod problem when no slope limiter is applied. We obtain a much sharper shock, but accentuate some of the oscillations visible around the head and tail of the rarefaction and ahead of the shock. With a slope limiter (Fig.~\ref{fig:sodrecon2}) the solution is similar to Fig~\ref{fig:sodstd} but with a much sharper shock front. In both cases the linear wave solution shows essentially no amplitude error over the 10 periods simulated (Figs.~\ref{fig:waverecon0} and \ref{fig:waverecon2}).
 
\section{Conclusion}
 I propose a minimal conversion of the SPH artificial viscosity into a `Godunov-type scheme' where we reconstruct the velocity field between particles to eliminate unwanted dissipation in smooth flow. This avoids the need for switches with arbitrary parameters and gives sharper shocks. While we presented 1D tests, preliminary findings in 3D are promising.

\section*{Acknowledgment}
% optional entry into table of contents (if used)
%\addcontentsline{toc}{section}{Acknowledgment}
 This is new material, and I would like to thank Matthew Bate for encouragement to write this up and the organisers for the opportunity to present my otherwise unpublished thoughts.



% http://www.ctan.org/tex-archive/macros/latex/contrib/supported/IEEEtran/bibtex
\bibliographystyle{IEEEtran.bst}
\bibliography{dan}
%\begin{thebibliography}{1}


%\end{thebibliography}

\end{document}